\documentclass[11pt]{article}
\usepackage{Blois,epsfig}

\begin{document}


\title{Novel Aspects of Hard Diffraction in QCD}

\author{Stanley J. Brodsky}

\address{Stanford Linear Accelerator Center, Stanford University, Stanford, CA, 94309}

\maketitle\abstracts{Initial- and final-state interactions from
gluon-exchange, normally neglected in the parton model have a
profound effect in QCD hard-scattering reactions, leading to
leading-twist single-spin asymmetries, diffractive deep inelastic
scattering, diffractive hard hadronic reactions, and nuclear
shadowing and antishadowing---leading-twist  physics not
incorporated in the light-front wavefunctions of the target computed
in isolation.  I also discuss the use of diffraction to materialize
the Fock states of a hadronic projectile and test QCD color
transparency.}

\section{Diffractive Deep Inelastic Scattering}

A remarkable feature of deep inelastic lepton-proton scattering at
HERA is that approximately 10\% events are
diffractive~\cite{Adloff:1997sc,Breitweg:1998gc}: the target proton
remains intact, and there is a large rapidity gap between the proton
and the other hadrons in the final state.  These diffractive deep
inelastic scattering (DDIS) events can be understood most simply
from the perspective of the color-dipole model: the $q \bar q$ Fock
state of the high-energy virtual photon diffractively dissociates
into a diffractive dijet system.  The exchange of multiple gluons
between  the color dipole of the $q \bar q$ and the quarks of the
target proton neutralizes the color separation and leads to the
diffractive final state.  The same multiple gluon exchange also
controls diffractive vector meson electroproduction at large photon
virtuality.\cite{Brodsky:1994kf}  This observation presents a
paradox: if one chooses the conventional parton model frame where
the photon light-front momentum is negative $q+ = q^0 + q^z  < 0$,
the virtual photon interacts with a quark constituent with
light-cone momentum fraction $x = {k^+/p^+} = x_{bj}.$  Furthermore,
the gauge link associated with the struck quark (the  Wilson line)
becomes unity in light-cone gauge $A^+=0$. Thus the struck
``current" quark apparently experiences no final-state interactions.
Since the light-front wavefunctions $\psi_n(x_i,k_{\perp i})$ of a
stable hadron are real, it appears impossible to generate the
required imaginary phase associated with pomeron exchange, let alone
large rapidity gaps.

This paradox was resolved by Paul Hoyer, Nils Marchal, Stephane
Peigne, Francesco Sannino and myself.\cite{Brodsky:2002ue} Consider
the case where the virtual photon interacts with a strange
quark---the $s \bar s$ pair is assumed to be produced in the target
by gluon splitting.  In the case of Feynman gauge, the struck $s$
quark continues to interact in the final state via gluon exchange as
described by the Wilson line.  The final-state interactions occur at
a light-cone time $\Delta\tau \simeq 1/\nu$ shortly after the
virtual photon interacts with the struck quark. When one integrates
over the nearly-on-shell intermediate state, the amplitude acquires
an imaginary part. Thus the rescattering of the quark produces a
separated color-singlet $s \bar s$ and an imaginary phase. In the
case of the light-cone gauge $A^+ = \eta \cdot A =0$, one must also
consider the final-state interactions of the (unstruck) $\bar s$
quark. The gluon propagator  in light-cone gauge $d_{LC}^{\mu\nu}(k)
= (i/k^2+ i \epsilon)\left[-g^{\mu\nu}+\left(\eta^\mu k^\nu+
k^\mu\eta^\nu / \eta\cdot k\right)\right] $ is singular at $k^+ =
\eta\cdot k = 0.$ The momentum of the exchanged gluon $k^+$ is of ${
\cal O}{(1/\nu)}$; thus rescattering contributes at leading twist
even in light-cone gauge.   The net result is  gauge invariant and
is identical to the color dipole model calculation. The calculation
of the rescattering effects on DIS in Feynman and light-cone gauge
through three loops is given in detail for an Abelian model in the
references.~\cite{Brodsky:2002ue}    The result shows that the
rescattering corrections reduce the magnitude of the DIS cross
section in analogy to nuclear shadowing.

A new understanding of the role of final-state interactions in deep
inelastic scattering has thus emerged. The multiple  scattering of
the struck parton via instantaneous interactions in the target
generates dominantly imaginary diffractive amplitudes, giving rise
to an effective ``hard pomeron" exchange.  The presence of a
rapidity gap between the target and diffractive system requires that
the target remnant emerges in a color-singlet state; this is made
possible in any gauge by the soft rescattering.  The resulting
diffractive contributions leave the target intact  and do not
resolve its quark structure; thus there are contributions to the DIS
structure functions which cannot be interpreted as parton
probabilities~\cite{Brodsky:2002ue}; the leading-twist contribution
to DIS  from rescattering of a quark in the target is a coherent
effect which is not included in the light-front wave functions
computed in isolation. One can augment the light-front wave
functions with a gauge link corresponding to an external field
created by the virtual photon $q \bar q$ pair
current.\cite{Belitsky:2002sm,Collins:2004nx}   Such a gauge link is
process dependent~\cite{Collins:2002kn}, so the resulting augmented
LFWFs are not
universal.\cite{Brodsky:2002ue,Belitsky:2002sm,Collins:2003fm}   We
also note that the shadowing of nuclear structure functions is due
to the destructive interference between multi-nucleon amplitudes
involving diffractive DIS and on-shell intermediate states with a
complex phase. In contrast, the wave function of a stable target is
strictly real since it does not have on-energy-shell intermediate
state configurations.   The physics of rescattering and shadowing is
thus not included in the nuclear light-front wave functions, and a
probabilistic interpretation of the nuclear DIS cross section is
precluded.

Rikard Enberg, Paul Hoyer, Gunnar Ingelman and
I~\cite{Brodsky:2004hi} have shown that the quark structure function
of the effective hard pomeron has the same form as the quark
contribution of the gluon structure function. The hard pomeron is
not an intrinsic part of the proton; rather it must be considered as
a dynamical effect of the lepton-proton interaction. Our QCD-based
picture also applies to diffraction in hadron-initiated processes.
The rescattering is different in virtual photon- and hadron-induced
processes due to the different color environment, which accounts for
the  observed non-universality of diffractive parton distributions.
This framework also provides a theoretical basis for the
phenomenologically successful Soft Color Interaction (SCI)
model~\cite{Edin:1995gi} which includes rescattering effects and
thus generates a variety of final states with rapidity gaps.

\section{ Single-Spin Asymmetries from Final-State
Interactions}

Among the most interesting polarization effects are single-spin
azimuthal asymmetries  in semi-inclusive deep inelastic scattering,
representing the correlation of the spin of the proton target and
the virtual photon to hadron production plane: $\vec S_p \cdot \vec
q \times \vec p_H$.  Such asymmetries are time-reversal odd, but
they can arise in QCD through phase differences in different spin
amplitudes. In fact, final-state interactions from gluon exchange
between the outgoing quarks and the target spectator system lead to
single-spin asymmetries in semi-inclusive deep inelastic
lepton-proton scattering  which  are not power-law suppressed at
large photon virtuality $Q^2$ at fixed
$x_{bj}$~\cite{Brodsky:2002cx} In contrast to the SSAs arising from
transversity and the Collins fragmentation function, the
fragmentation of the quark into hadrons is not necessary; one
predicts a correlation with the production plane of the quark jet
itself. Physically, the final-state interaction phase arises as the
infrared-finite difference of QCD Coulomb phases for hadron wave
functions with differing orbital angular momentum.  The same proton
matrix element which determines the spin-orbit correlation $\vec S
\cdot \vec L$  also produces the anomalous magnetic moment of the
proton, the Pauli form factor, and the generalized parton
distribution $E$ which is measured in deeply virtual Compton
scattering. Thus the contribution of each quark current to the SSA
is proportional to the contribution $\kappa_{q/p}$ of that quark to
the proton target's anomalous magnetic moment $\kappa_p = \sum_q e_q
\kappa_{q/p}$.\cite{Brodsky:2002cx,Burkardt:2004vm}  The HERMES
collaboration has recently measured the SSA in pion
electroproduction using transverse target
polarization.\cite{Airapetian:2004tw}  The Sivers and Collins
effects can be separated using planar correlations; both
contributions are observed to contribute, with values not in
disagreement with theory
expectations.\cite{Airapetian:2004tw,Avakian:2004qt} A related
analysis also predicts that the initial-state interactions from
gluon exchange between the incoming quark and the target spectator
system lead to leading-twist single-spin asymmetries in the
Drell-Yan process $H_1 H_2^\updownarrow \to \ell^+ \ell^- X$.
\cite{Collins:2002kn,Brodsky:2002rv}    The SSA in the Drell-Yan
process is the same as that obtained in SIDIS, with the appropriate
identification of variables, but with the opposite sign.
Initial-state interactions also lead to a $\cos 2 \phi$ planar
correlation in unpolarized Drell-Yan reactions.\cite{Boer:2002ju}
There is no Sivers effect in charged-current reactions since the $W$
only couples to left-handed quarks.\cite{Brodsky:2002pr}

\section{Diffraction Dissociation as a Tool to Resolve Hadron Substructure}

Diffractive multi-jet production in heavy nuclei provides a novel
way to resolve the shape of light-front Fock state wave functions
and test color transparency.\cite{Brodsky:1988xz}   For example,
consider the reaction~\cite{Bertsch:1981py,Frankfurt:1999tq} $\pi A
\rightarrow {\rm Jet}_1 + {\rm Jet}_2 + A^\prime$ at high energy
where the nucleus $A^\prime$ is left intact in its ground state. The
transverse momenta of the jets balance so that $ \vec k_{\perp i} +
\vec k_{\perp 2} = \vec q_\perp < {R^{-1}}_A \ . $ Because of color
transparency, the valence wave function of the pion with small
impact separation will penetrate the nucleus with minimal
interactions, diffracting into jet pairs.\cite{Bertsch:1981py}  The
$x_1=x$, $x_2=1-x$ dependence of the di-jet distributions will thus
reflect the shape of the pion valence light-cone wave function in
$x$; similarly, the $\vec k_{\perp 1}- \vec k_{\perp 2}$ relative
transverse momenta of the jets gives key information on the second
transverse momentum derivative of the underlying shape of the
valence pion wavefunction.\cite{Frankfurt:1999tq,Nikolaev:2000sh}
The diffractive nuclear amplitude extrapolated to $t = 0$ should be
linear in nuclear number $A$ if color transparency is correct.  The
integrated diffractive rate will then scale as $A^2/R^2_A \sim
A^{4/3}.$ This is in fact what has been observed by the E791
collaboration at FermiLab for 500 GeV incident pions on nuclear
targets.\cite{Aitala:2000hc}   The measured momentum fraction
distribution of the jets is found to be approximately consistent
with the shape of the pion asymptotic distribution
amplitude.\cite{Lepage:1979zb,Efremov:1978rn,Lepage:1980fj}
$\phi^{\rm asympt}_\pi (x) = \sqrt 3 f_\pi
x(1-x)$.\cite{Aitala:2000hb}  Remarkably this is also the prediction
of AdS/CFT duality for the light-front wavefunctions of the pion in
conformal QCD.\cite{Brodsky:2003px}

The concept of high energy diffractive dissociation can be
generalized to provide a tool to materialize the individual Fock
states of a hadron, photon, or nuclear projectile; {\em e.g.}, the
diffractive or Coulomb dissociation of a high energy proton $p A \to
qqq A^\prime$  or $p e \to qqq e$ can be used to measure the valence
light-front wavefunction of the proton as well as its intrinsic
heavy quark Fock states. Similarly, the hidden-color Fock
states~\cite{Brodsky:1983vf} of the six-quark deuteron, can be
dissociated to final states such as $\Delta^{++} \Delta^{-}.$

\section{Antishadowing of Nuclear Structure Functions}

One of the novel features of QCD involving nuclei is the {\it
antishadowing} of the nuclear structure functions which is observed
in deep inelastic lepton scattering and other hard processes.
Empirically, one finds $R_A(x,Q^2) \equiv  \left(F_{2A}(x,Q^2)/
(A/2) F_{d}(x,Q^2)\right) > 1 $ in the domain $0.1 < x < 0.2$; {\em
i.e.}, the measured nuclear structure function (referenced to the
deuteron) is larger than the scattering on a set of $A$ independent
nucleons. The shadowing of the nuclear structure functions:
$R_A(x,Q^2) < 1 $ at small $x < 0.1 $ can be readily understood in
terms of the Gribov-Glauber theory.  Consider the two-step process
illustrated in Fig.  1 in the nuclear target rest frame. The
incoming $q \bar q$ dipole first interacts diffractively $\gamma^*
N_1 \to (q \bar q) N_1$ on nucleon $N_1$ leaving it intact.  This is
the leading-twist diffractive deep inelastic scattering  (DDIS)
process which has been measured at HERA to constitute approximately
10\% of the DIS cross section at high energies.  The $q \bar q$
state then interacts inelastically on a downstream nucleon $N_2:$
$(q \bar q) N_2 \to X$. The phase of the pomeron-dominated DDIS
amplitude is close to imaginary, and the Glauber cut provides
another phase $i$, so that the two-step process has opposite  phase
and  destructively interferes with the one-step DIS process $\gamma*
N_2 \to X$ where $N_1$ acts as an unscattered spectator. The
one-step and-two step amplitudes can coherently interfere as long as
the momentum transfer to the nucleon $N_1$ is sufficiently small
that it remains in the nuclear target;  {\em i.e.}, the Ioffe
length~\cite{Ioffe:1969kf} $L_I = { 2 M \nu/ Q^2} $ is large
compared to the inter-nucleon separation. In effect, the flux
reaching the interior nucleons is diminished, thus reducing the
number of effective nucleons and $R_A(x,Q^2) < 1.$

\begin{figure}
\centering \psfig{figure=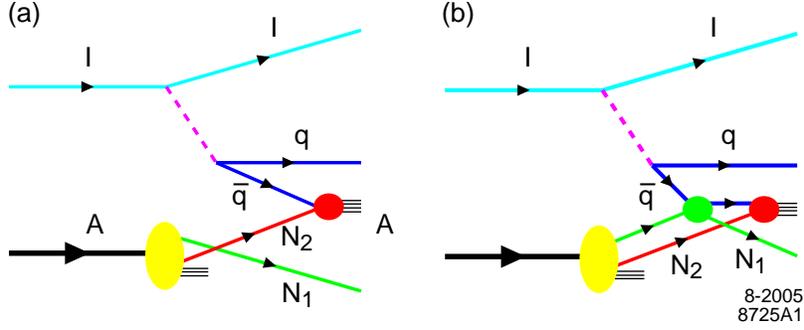,height=1.75in}
\caption{Illustration of one-step and two-step processes.
\label{fig:radisha}}
\end{figure}

There are also leading-twist diffractive contributions $\gamma^* N_1
\to (q \bar q) N_1$  arising from Reggeon exchanges in the
$t$-channel.\cite{Brodsky:1989qz}  For example, isospin--non-singlet
$C=+$ Reggeons contribute to the difference of proton and neutron
structure functions, giving the characteristic Kuti-Weisskopf
$F_{2p} - F_{2n} \sim x^{1-\alpha_R(0)} \sim x^{0.5}$ behavior at
small $x$. The $x$ dependence of the structure functions reflects
the Regge behavior $\nu^{\alpha_R(0)} $ of the virtual Compton
amplitude at fixed $Q^2$ and $t=0.$ The phase of the diffractive
amplitude is determined by analyticity and crossing to be
proportional to $-1+ i$ for $\alpha_R=0.5,$ which together with the
phase from the Glauber cut, leads to {\it constructive} interference
of the diffractive and nondiffractive multi-step nuclear amplitudes.
Furthermore, because of its $x$ dependence, the nuclear structure
function is enhanced precisely in the domain $0.1 < x <0.2$ where
antishadowing is empirically observed.  The strength of the Reggeon
amplitudes is fixed by the fits to the nucleon structure functions,
so there is little model dependence.

As noted in Section 1, the Bjorken-scaling diffractive contribution
to DIS arises from the rescattering of the struck quark after it is
struck  (in the  parton model frame $q^+ \le 0$), an effect induced
by the Wilson line connecting the currents. Thus one cannot
attribute DDIS to the physics of the target nucleon computed in
isolation.\cite{Brodsky:2002ue}  Similarly, since shadowing and
antishadowing arise from the physics of diffraction, we cannot
attribute these phenomena to the structure of the nucleus itself:
shadowing and antishadowing arise because of the $\gamma^* A$
collision and the history of the $q \bar q$ dipole as it propagates
through the nucleus.

In a recent paper, Ivan Schmidt, Jian-Jun Yang, and
I~\cite{Brodsky:2004qa} have extended this analysis to the shadowing
and antishadowing of all of the electroweak structure functions.
Quarks of different flavors  will couple to different Reggeons; this
leads to the remarkable prediction that nuclear antishadowing is not
universal; it depends on the quantum numbers of the struck quark.
This picture leads to substantially different antishadowing for
charged and neutral current reactions, thus affecting the extraction
of the weak-mixing angle $\theta_W$.  See Fig. 2. We find that part
of the anomalous NuTeV result~\cite{Zeller:2001hh} for $\theta_W$
could be due to the non-universality of nuclear antishadowing for
charged and neutral currents. Detailed measurements of the nuclear
dependence of individual quark structure functions are thus needed
to establish the distinctive phenomenology of shadowing and
antishadowing and to make the NuTeV results definitive. Schmidt,
Yang, and I have also identified contributions to the nuclear
multi-step reactions which arise from odderon exchange and also
hidden color degrees of freedom in the nuclear wavefunction. There
are other ways in which this new view of antishadowing can be
tested;  antishadowing can also depend on the target and beam
polarization.

\begin{figure}[htb]
\centering\psfig{figure=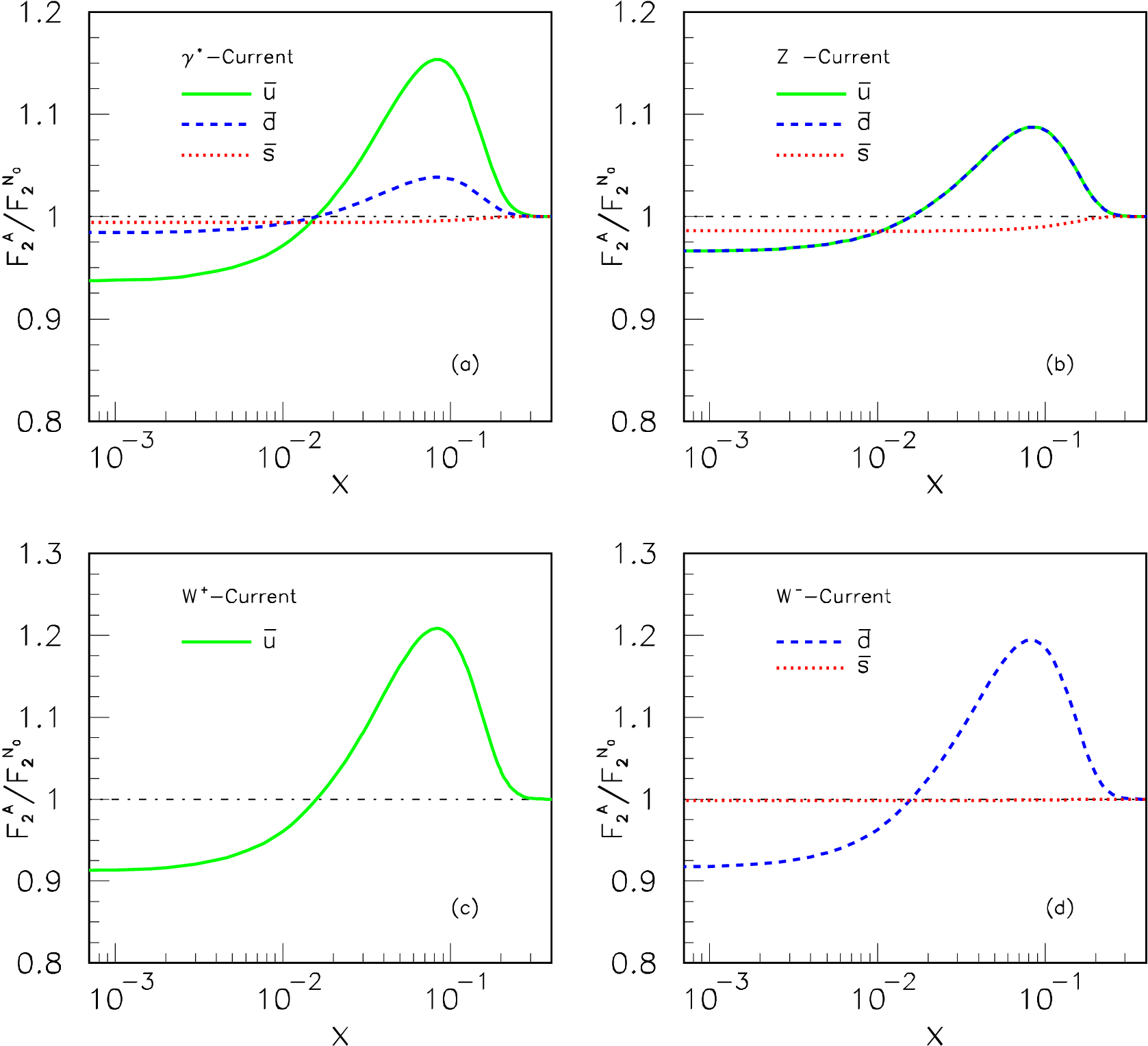,height=4.5in}
\caption[*]{Model predictions~\cite{Brodsky:2004qa} for interactions
of electroweak interactions on antiquarks in nuclear targets. The
antishadowing effect is not as large for quark currents.
\label{fig:radish}}
\end{figure}

\section*{Acknowledgments}
This
talk is based on collaborations with Rikard Enberg, Paul Hoyer, Dae Sung Hwang,
Gunnar Ingelman,  Hung Jung Lu, Ivan Schmidt and Jian-Jun Yang. The work was supported in
part by the Department of Energy, contract No. DE-AC02-76SF00515.

\end{document}